# Crawling political communities in Twitter and extracting political affiliations

**Muhammad Umer Gurchani**

**Université de Montpellier**

**muhammad-umer.gurchani@etu.umontpellier.fr**

## Abstract :

In theory, a major advantage to the big data approach in studying online communities is that it should be possible to collect a representative random sample from a broadly defined population. However, in practice, data collection processes are not formalized, even for famous social media platforms such as Twitter and Facebook. As a result, there is ambiguity left on questions like 'how much data is enough?' and how representative are the samples of the broader population being studied in online social networks.

In this Paper, I propose a focused back-and-forth crawl approach and a validated seed choice method for collecting network level data from Twitter. The proposed crawl method can extract community structures without needing a complete network graph for the Twitter network and validate it's size using 'reference score' (Blenn, et al. 2012). It also takes care of the sampling size problem in Twitter by tracking the percentage of known nodes that have been included in the data. Thus, solving most major problems in Twitter data collection procedures and moving a step further to formalizing data collection methods for the platform.

Once the communities are crawled, and the network graph is clean and complete; it is then possible to train Machine Learning classifiers using communities as features to predict political affiliations of users on a larger scale. As a case, I used the proposed method for separating French political communities on Twitter from the global Twitter community and knowing political affiliations of users on a continuous scale.

### Introduction

As the popularity of Twitter is growing, it is becoming increasingly hard for social scientists to gather profile samples that represent the target population in a study. The standard method of gathering samples[1] for research on questions such as 'Is Twitter Polarizing its user?', includes arbitrary cut-off points which make the data collection process more convenient but less representative (Blenn, et al. 2012). As there are more and more researchers using Twitter for studying questions in social sciences, there is a need to formalize the data gathering process while keeping in mind that Twitter's network data comes in graph format which has its additional traversal and sampling complications that need to be addressed (Wang, et al. 2011).

One of the major problems of using the network topology of social networks for answering sociological questions is that sampling methods for large network graphs is still an open question (Krishnamurthy, et al. 2007) and any sort of missing data can have a serious effect on results (Smith, Moody et Morgan 2017) (Zhang et Patone 2017) (Zhang, et al. 2015). The Network-graph sampling problem is different from traditional sampling problems where snow-ball sampling (or other forms of convenient sampling) is used. Where snow-ball sampling can help know characteristics of nodes, it tells nothing about the characteristics of the overall network such as degree distribution (Ribeiro et Towsley 2012), betweenness centrality distribution, average path length, assortativity, and clustering coefficient (Wu, et al. s.d.) and modularity, which are important measures for investigations on questions on network homophily and polarization. For such questions, it is important to devise sampling methods that reflect the characteristics of the network (Lee, Kim et Jeong 2006).

The minimum sample size of a network that could preserve its topological properties is also an important and relevant debate when it comes to studying a large social network. To the best of my knowledge, even the smallest proposed sample size that accurately represents the characteristics of the full graph constitutes at least 15 percent of the original population size (Leskovec et Faloutsos 2006). Jure Leskovec's method does well for finding the sample of a known graph but in most real-world scenarios larger network graphs are unavailable and their sizes are unknown. For example, if a researcher is trying to answer a sociological question using Twitter's network data from a particular country, it Is difficult to know if the sample of profiles he/she is using represents the rest of the community since Twitter does not tell us how many profiles of a particular community there are in total. Most studies in political science and sociology that use data from social media, tend to ignore this problem and use an 'arbitrary' number of profiles who tweeted a particular hashtag to reach conclusions on questions such as the extent of political polarization and homophily (for which graph modularity is an important measure). While such inquiries can help establish the extent of polarization on an issue (being discussed through hashtag), long-term network-based polarization which is known to be caused by higher levels of homophily requires that we investigate network graphs and relationships as-well (Cass R 1999). Thus, graph sampling problem requires more attention and inquiries from social scientists.

---

[1] By 'Standard method', I mean methods that use stream-api to collect live tweets and process them to get users.

In this Paper, I argue that graph-sampling problem becomes more complicated when the population Graph is unknown (which is usually the case in Twitter investigations). I then suggest a comprehensive method for seed selection and community crawl that ensures that at least 50 percent (in the worst-case scenario) of the target community is included in the study which is enough to act as a target population. Only then, it becomes possible to collect a representative sample of the target population and study the community structure and try to answer sociological questions that require network topology. I go further by extracting information about features such as political affiliations from the crawled network graph and show that high level of accuracy can be reached with machine learning models when predicting political affiliation using relationships in Twitter.

Most of the work concerned with political polarization in Twitter has been using data from stream API which only provides information on nodes that are active on a particular political issue (represented by a hashtag) (Placeholder1). While this type of studies can accurately estimate issue-based polarization, but they fail to capture the long-term picture of Twitter network. Due to close relationship between political homophily and polarization, it is possible to treat polarization as a latent variable (Placeholder2) dependent on levels of homophily which have been studied using 'graph modularity' in Twitter (Placeholder3). Although, It is not unprecedented to study political polarization and homophily in this manner but in recent year, it has become a less common to use modularity as an indicator of homophily Twitter as it has grown into a highly diverse network in many countries and contains a lot of different type of communities that range from science, fashion, sports and politics. With such diversity in location and interests among users, modularity does not make sense as an indicator of homophily as people can be in the same community due to shared interest or location. To use modularity score as an indicator of pollical homophily, it is important to separate a political network belonging to a single political context from the global network. It is thus high time to formalize crawling mechanisms in Twitter so that it is possible to validate the representativeness of the data collected through crawls. Since the crawls are highly targeted, we will only get users from a single political context, it will be meaningful to use modularity score as an indicator or political homophily. In the case of this paper, it is assumed that French National politics falls in a single political context and methods used for studying homophily in French network can be replicated for other national political contexts in Twitter

## Review of Graph Traversal Methods:

### Breadth-First Crawl:

Breadth-First is the most widely used crawl mechanism in network graphs. BFS starts with a list of seed nodes and crawls all the neighbors of each node included in the list while creating a new list of found nodes. Once all the neighbors of each seed node have been explored, it moves to the newly found nodes and repeats the same process. The time complexity of BFS is $O(|V|+|E|)$. The selection of seeds is of paramount importance in BFS if crawling a subgraph is an objective (which is the case in this research).

### Depth-First Crawl:

"Depth-First algorithm starts at the root node (selecting some arbitrary node as the root node in the case of a graph) and explores as far as possible along each branch before backtracking.". For problems like crawling of Twitter, DF is highly impractical as many nodes are connected to people outside of the target community and it will be very easy to lead the crawler out of the target nodes.

### Random-Walk:

Random walks are by far the most widely used method for both traversing and sampling network graphs. It starts with a randomly selected node and 'selects the next node at random from the neighbors'. While Random walks can represent the graph properties very well but they can not be used on Twitter for inquiries on questions of homophily unless one has access to the target population. Since questions about homophily are usually focused towards a political context (like French Politics or American Politics) and size of community that Is part of that context is unknown in Twitter, it will be impossible to know when to stop the crawl and what percentage of target community has been crawled (I will solve this problem using a variation of BFS in this Paper)

## Review on Graph Sampling:

Over the last few years, there have been developments in graph sampling problems in general but not a lot of attention has been given specifically to Twitter's network sampling issues (Ahmad, et al. 2010). Here are a few sampling methods generally used for sampling large graphs.

**Random Node Sampling:**

Random Node Sampling is a highly used technique that works well to represent topological properties of a network graph only if, the larger graph is known, and samples are at least 15 percent of the complete network (Leskovec et Faloutsos 2006). Due to it's large size and unavailability of complete data, the Twitter network graph can not be represented using Random-Node Sampling.

**Random Degree Node Sampling:**

Random Degree node sampling is a variant of random node selection but only adjusts the probability of being selected in the sample using degrees of nodes.

**Random-Edge Sampling:**

Random-Edge sampling is another technique used for sampling relationships. It randomly selects the edges in a graph and represents it a sample. Its ability to represent properties of the network has been tested in different studies and this method has proved ineffective as a sampling method for questions where network properties are needed (Wu, et al. s.d.).

## Problem description

In a well-connected global social network like Twitter, it is now more likely than ever for a national politician to be globally popular. This global Twitter network can be formally interpreted as a graph with nodes acting as users and edges acting as relationships. We can call this graph, $G(V, E)$, where $V$ is a set of vertices (profiles of individuals) and $E$ is a set of $V$'s edges (follow relationship between individual). Within this graph we know that there exist several sub-graphs where nodes are highly connected to each-other but less connected to other sub-graphs. Since our goal is to measure group polarization in French context, we are looking to capture only one sub-graph (French Political Community) and separate it from other sub-graphs in the global network so that we can measure political polarization with in it. This French graph is entrenched in the global network and overlaps many other sub-graphs of the global network for example many French users tend to follow American, British and German politicians and celebrities with millions of followers. If we do not separate the French subgraph, then our automated crawler will end up crawling millions of International users who will have very little interest in French politics and will act as noise in measurement of group polarization phenomena.

For clarity we will call the French political subgraph as f(v,e). In order to get a rough idea of audience I will use statistics published for marketing purposes by multiple advertisers[3]. According to these numbers, the number of profiles V in G(V, E) stand at about 330 million whereas number of 'v' in f(v,e) stand around 2.5 million. Our objective here is to come up with a crawl strategy that only crawls f(v,e) users.

## Selecting Seed Profiles:

Goal of the crawl is especially important when It comes to choosing the best seed profiles, as they determine the starting point of the crawl. In our case, it is equally important to know the distribution of kind of users in the community structure that we end up detection. I will be looking at the following qualities in the seed profiles.

1. Their political affiliation must be known.
2. Seed profiles should be diverse enough to cover the political spectrum.
3. Political affiliation in profiles should be evenly distributed (There should be same number of profiles for each of the parties or political groups)

For France I will be using Twitter data from French Presidential elections of 2017 and will be using data of supporters of top 5 presidential candidates this election to create a seed profile dataset.

1. Emmanuel Macron
2. Marine Le Pen
3. Jean Luc Melenchon
4. Francois Fillon
5. Benoit Hammon

During the first round of presidential elections these candidates managed to gain over 91 percent of total voter turn-out which testifies to the fact that taking data of their supporters is likely to cover a large majority of French Political context in Twitter. In addition to that diversity of opinions of these candidates (Extreme Left to Extreme Right) also provides enough reason to believe that taking data of their supporters will serve as a good starting point to crawl.

A popular strategy for seed selection in Twitter research is to manually compile a list of profiles that are most popular in multiple communities.  While such a strategy can be effective, but it is difficult to test its validity. As mentioned above, in BFS based crawling algorithms seed selection is of paramount importance. Keeping in mind that the goal of the crawl is to maximize the number of profiles from the French Political context in Twitter, it is important to formalize the problem and come up with rules that can help with seed selection for Twitter in future research. In case of seed selection following questions are the most important ones to address.

1. How big should the seed database be?
2. What type of profiles should be included in the seed database?

To answer the above questions, I trained an AI model to approximate the best possible seed profiles using a manually annotated database collected during presidential elections of France in 2017 (Fraisier, et al. 2018). The main reason for using that dataset was that the political affiliation of users in that dataset was known (self-declared and manually verified). However, one limitation of this dataset was that it only marked the top 5 parties in France. Although these parties combined gained more than 91 percent of total votes cast in the first round of the 2017 presidential elections, but not having smaller parties in the seed selection would mean that the final database might not include smaller and more isolate communities (if there are any).

The first step towards the process was that I collected all the accounts followed by accounts included in Frasier's database ('Friends' in Twitter's official language). Rate-limit in Twitter made it exceedingly difficult to collect this data but parallel crawling using multiple Aps increase the speed of this process to some extent. This data can be expressed in form of sets.

Set $U$ represents the accounts in manually annotated database.

$$U = \{u_1, u_2, u_3 \cdots u_{22853}\}$$

Set $U_{friends}$ contains five different sets where sub-text indicate the party to which the set belongs. $F_{fn}$ which is one of the members of $U_{friends}$, contains sets of 'friends' of accounts from $U$ who support *Front National*

$$U_{friends} = \{F_{fn}, F_{ps}, F_{em}, F_{fi}, F_{lr}\}$$

Expressing the database in the above two formats, I am now in better position to formalize the seed selection problem. Since the goal of the crawl is to separate the profiles that are included in the French Political context from the global Twitter graph, seed profiles should be the most popular exclusive profiles within the target community. This problem can be mathematically expressed as:

From $\cup F_{fn}$ select minimum number of elements to form a set $S_{fn}$ (seed profiles for FN) such that set $S_{fn}$ meets the following conditions:

**Condition 1** : $|S_{fn} \cap (\cup F_{fn})|$ is maximized

**Condition 2**: $|S_{fn} \cap (\cup(U_{friends} - \cup F_{fn}))|$ is minimized

This is multi-objective optimization problem with no exact solution, and I will now try to assess it's difficulty and propose a best course of action for our case. First condition of above problem (as stated) is a classical problem in combinatorics, 'set hitting problem' and has been categorized as NP-complete problem which makes it computationally expensive to come up with best possible solution (Fijany et Vatan 2004). Approximation is the usual course in set-hitting problem if we are dealing with large datasets.

The second condition was added after testing the first one as the resulting profiles were out of context such as those of famous footballer, Actors and American celebrities. By minimizing $|S \cap (\cup(U_{friends} - \cup F_{fn}))|$ it is ensured that we will only get profiles that are exclusively popular in one political party. At the seed stage it will suffice to get exclusive profiles. Choosing universally popular profiles is not an optimum approach as they are bound to be part of the graph in the third step of this crawl and including them at this stage will only diverge the crawl outside of target population (French political context).

I approached this problem by solving it in two steps to find a final approximate solution. In the first step, I will use the following greedy algorithm to fulfill the first condition. Once I have enough profiles to cover data that meets the first condition, I will then take profiles from that set which will also meet the second criteria.

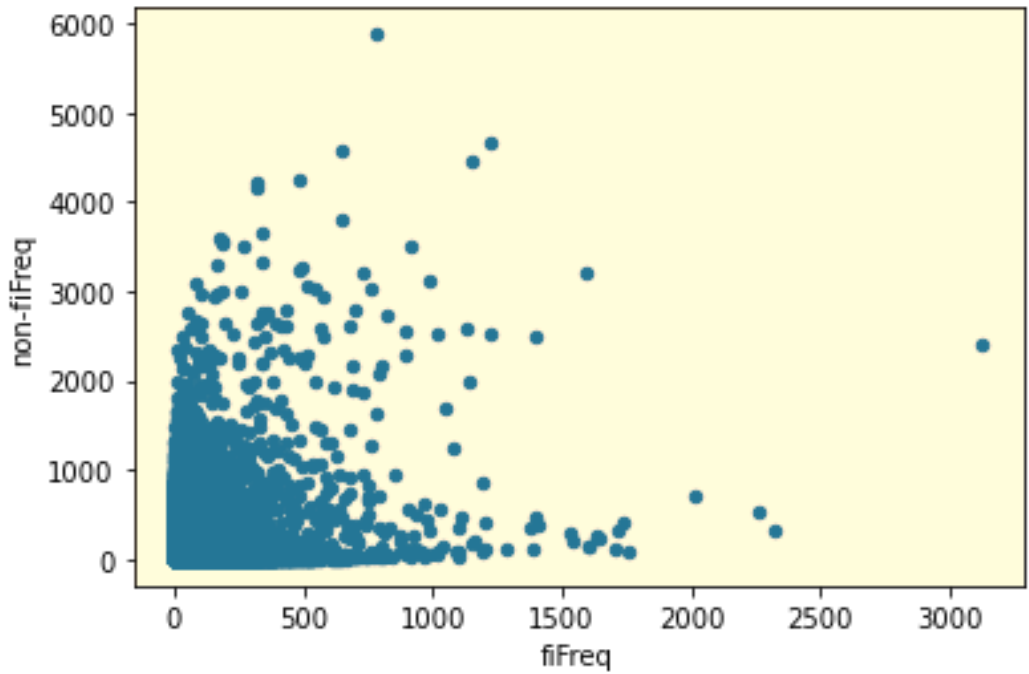

**Fig 1:** This is the plot of most popular accounts in France Insoumis according to manually annotate database. On x-axis, we have their frequency of being followed by France Insoumis and on y-axis by other parties combined. Best seed profiles for France Insoumis lie in the botton right corner.

Following greedy approach was used to come up with a solution that fulfills the first condition.

**Algorithm 1: Greedy Algorithm for set-hitting problem**

**Input: Set of sets U, k**

**2 for i = 1...k do**

       **3 pick the set that covers the maximum number of uncovered elements**

It was observed that, just to get all over 5000 *France Insoumis* profiles from manually annotated database, I will need at least 502 seed profiles. By that observation, it was inferred that if cost and time of crawling were not a factor, over 2500 seed profiles, would be needed to get a comprehensive database of French profiles in Twitter for all major political parties. In order to reduce the time of crawl, a compromise was reached using the observation that 84.5 percent of the profiles in the manually annotated database that support FI could be covered using under 30 seed profiles. Similar compromises were made for all parties and it was ensured that seed profiles are selected that would cover at least over 80 percent of the manually annotated database. Using the above method, a final dataset of 122 seed profiles was compiled.

To make sure that seed profiles are exclusively popular in only one of the political parties, profiles that returned a value of less than 0.8 for the ratio $\frac{\# \, of \, Followers \, from \, one \, party}{\# \, Total \, Followers \, in \, Database}$ were taken out of the seed database.

<u>**Resulting Seed Profiles**</u>

| Party | Number of Seed Profiles | Percentage of manual database profiles covered[3] | Types of Seed Profiles found |
|---|---|---|---|
| *France Insoumise* | <u>**25**</u> | **85.16** | Second tier party leadership, Allies in 2017, party activists, Official and unofficial party accounts |
| *En Marche* | <u>**33**</u> | <u>**86.6**</u> | Second tier party leadership, Former *Parti Socialiste* and EM activists |
| *Front National* | 16 | 84.5 % | Second tier party leadership, party accounts, Anonymous Accounts, party activists |
| *Parti Socialiste* | **55** | **68.4 %** | Former and current party leader, Official and unofficial party accounts |
| *Les Republicains* | 26 | 91.226 % | Party Leadership, Official and unofficial party accounts |
| **Total** | | | |

# Crawl Algorithm

Knowing the trade-offs from each of the directions, I propose the following back and forth scheme between elites and ordinary users for crawling a complete political context in Twitter

1. Start with a database of known profiles with known political affiliation
2. Select the best seed profiles that fit best with known data
3. Towards Ordinary + shortlist
4. Towards Elite + shortlist
5. Move to step 3, unless the average Reference score is close to 50 percent (Justify).

**Advantages of the above crawling:**

Main advantage of the above crawling mechanism is that it will move from a targeted community (People embedded in French Political Context) and crawl that community first before moving to next one (at which point I stopped the algorithm)

**Limitations of above crawling**

Although the above proposed crawling mechanism will be effective in terms of its ability to crawl through political context and profiles involved in political discussion but the down-side this method is the run-time it might take to finish the crawl. Twitter API does not allow crawling without application of a rate limit which can prove to be a serious bottle-neck for time required for such crawls to finish. In order to overcome this limitation, parallel crawling on limited scale was applied which solved the problem to some extent but it still remains a significant issue that can limit the future studies.

Another limitation of the crawling mechanism proposed above is that, there is many profiles who are highly embedded in many different political contexts (such as both France and Algeria) but here we will work under the assumption that once the crawl is complete and we can run community detection algorithm on the detected profiles, we will be able to find the profiles that are from a completely different political context as they will be clustered together and manual analysis will eliminate that possibility.

**What determines the direction in each step?**

# How will I validate how much data is enough?

## Reference Score

To the best of my knowledge, the first paper that has tried to crawl communities in social media without using the complete graph came from *Blenn, Doerr, Kester, and Piet Van Mieghem*[5] in their seminal work published in 2012 which allowed for faster crawling of communities without knowing the full graph. They called their method "*Mutual Friend Crawling*" which was based on a strategy of crawling nodes that the highest "reference score" which they defined as following:

$$S_R = \frac{Found\ Referances\ of\ Each\ Node}{Degree\ of\ Each\ Node}$$

Major benefit of crawling with *Mutual Friend crawling* method was that it allowed to crawl densely connected communities first and then moved to next community that would be most connected to the first community. MFC approaches the question by crawling from a community towards the broader graph but fails to explain when to stop the crawl and does not clarify the different results that one might (or may not) obtain by choosing different starting point for the crawl. In this paper, I will address this question and try to define a well targeted community to crawl in a social network and suggest how Mutual friend crawling can be customized to crawl a political community on Twitter. I will also analyze the role seeds can play in crawling with community first approach.

## Setting a Target Reference Score:

In order to make sure that the crawler has been successful in getting a large portion of the targeted graph, it is important to set a goal reference score. This goal reference score can be known from the profiles that we know already (I.e: manually annotated profiles). Setting a concrete value for reference score requires that we manually look into the friends of a sample these profiles and for each profile we find out the ration of his/her friends who belong to the political context we are targeting to the total number of Friends that they have. Once we know target score of all the profiles in sample, we can take an average of this score and the set it as our crawl's target reference score. (50 percent)

Write the formula for this

## Crawl Direction for efficient implementation of MFC

When crawling a social network like Twitter from any set of profiles A, there will always be two directions a crawler can take.

1. Towards Followers (Profiles who follow A)
2. Towards Friends (Profiles which A follows)

Each of these directions have unique features, which must be accounted for before devising the direction in which a crawler should move.

### Towards Followers (Ordinary)

Starting with a set of Twitter profiles A, crawling towards the followers will provide the following:

a. Edges (connections) between profiles included in A.
b. Larger set of profiles which are likely to be more 'ordinary' than profiles in set A[4]

**Towards Friends (Elite)**

Starting with a set of Twitter profiles A, crawling towards Friends will provide the following:

a. Edges (connections) between profiles included in A.
b. Larger set of profiles which are likely to be less 'ordinary' than profiles in set A

**Shortlisting Thresholds**

After each crawl, there will be a need to clean the results as each crawl from ordinary users towards their friends will give global celebrities (which are not specific to French context) such as Barack Obama who has to this date 126 million followers. If the result of this crawl are not cleaned and profiles such as Barack Obama are kept in the index then the next crawl (which will be finding followers of this index) will become extremely costly and ineffective as it will spend most of it's time in populating the network of these large profile (which are very important globally but not specific to French context). To keep the crawl manageable a small trade-off is needed here in favor of efficacy as opposite to accuracy. To materialize this trade-off a metric is suggested called **Shortlisting Threshold.** In order to make to make it inside the index, a Friend profile has to meet the following criterion:

1) Be followed by atleast (seed ordianry /number of targeted politicians)

When crawling from the elite profiles to the ordinary profiles, the same problem can occur as there will be many foreign profiles that follow French celebrity politicians or large medias and in

---

[4] Twitter is a highly stratified network which act like a news media and reciprocity of connections is rare compared to social networks like Facebook (https://dl.acm.org/doi/abs/10.1145/1772690.1772751 ) .

order for these profiles to not be part of the index following **Shortlisting Threshold** willl be applied:

   2) Be follower of at least (seed elite / number of targeted politicians)

**Judging the distribution of cluster among political parties**

Once the context has been crawled, the next step will be to estimate the political affiliation of large number of users from the seed profiles whose political affiliations are already known. In order to perform this step, it is important to know that Twitter is known to have highly homophilic connections between users which can be used for determining political affiliations of users in large scale using an affiliation of smaller number of known users.

Using the network graph gained from the step above, I will run modularity-based community detection algorithm to find out the community structure. For a graph of this size, using Louvain's algorithm was most feasible step forward considering that it has a linear run-time (time complexity) and maximizes the modularity. This algorithm was run 50 times to ensure the consistency of the results and resulting average modularity score of was found to be 0.40 with 12 communities.

Once the community structure has been determined, the distribution of political affiliations in each cluster was initially estimated using the location of each of the seed profiles that were self-declared and manually annotated. In order to cross-validate the predicted community structure, and to eliminate irrelevant clusters, random samples were extracted from each of the cluster and studied manually.

In addition to cluster's distribution of political affiliations, the embeddedness score of each individual in network of particular kind of politician was calculated using the following function.

E (Support candidate 1) = (Reference score) * (1/avg path-length to known support candidate 1)

"Embeddedness" in network of a political party or a political personality is not a guarantee that the individual is supporting that candidate, but it can serve as measurement tool on continues scale to judge the level of political homophily in a network.

**Validation of Data Using Machine Learning**

Position of each of the profiles could be used as feature in a Machine Learning Classifier and results could determine how accurate political predictions are. Confidence of the ML classifier can be used as "embeddedness Score", contrary to the above solution.

**Crawl Results**

Using 22000 seed profiles from 2017 presidential elections that described their own political affiliations, three crawls proved to sufficient to achieve reference score close to 0.5. The first crawl was towards the elite, and resulted in 3000000 elite profiles but managed to achieve a poor reference score of 0.03 which is understandable considering that I had only explored one side of a bidirectional graph.

Decide how the features on second crawl were selected (mention Machine Learning or not) . The second crawl from the selected features towards their followers, fared much better in terms of reference score (0.17) which can be explained by the fact the most of the elite profiles do not have as many 'friends' as 'followers' in Twitter.

Third crawl was from ordinary profiles discovered in the second crawl towards the people who they follow. During this crawl, the numbers of input nodes were significantly larger than the previous two crawls and in total 1.7 million calls were made using parallel crawling techniques which resulted in 000000000000000000 million profiles. In order to reduce the graph to manageable level only the edges within the input graphs were kept and others were discarded. I also added profiles of elites discovered in the third crawl if they had significant following within the seed profiles (being followed by at least 10 people). This resulted in complete network graph of 2.2 million profiles with 35 million edges.

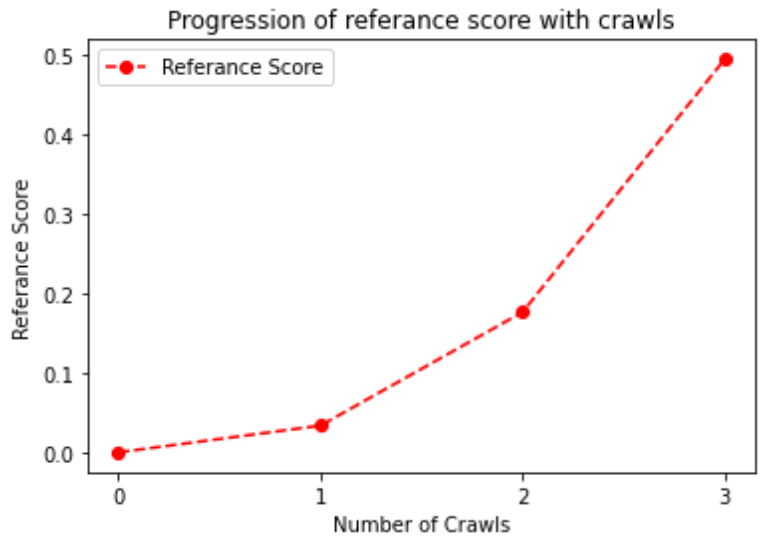

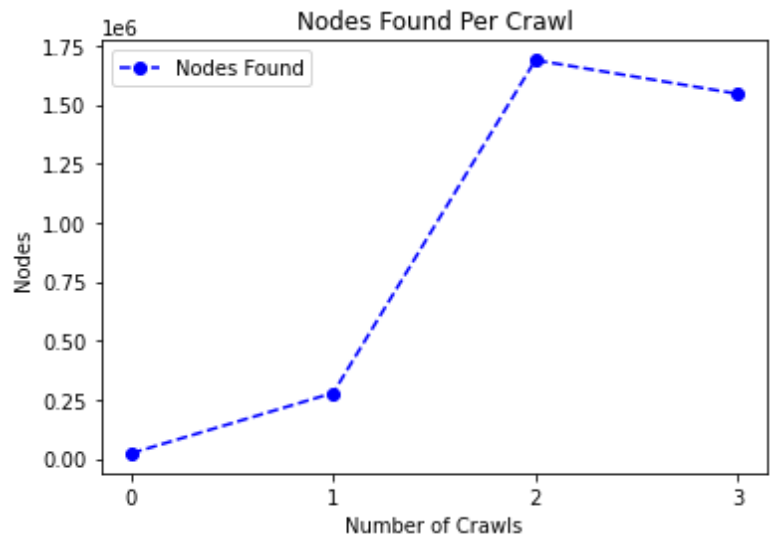

**Cluster Distibution Results:**

Once the crawls were complete and I had access to a decent proportion of the graph which is highly likely to retain the qualities of the complete graph (which can be double in size in worst case scenario as we are sure to have 50 percent of the data), it was ready to study for running a modularity-based clustering algorithm. Louvain's algorithm was chosen for the task due to its run-time in large network graphs. Running it on the complete graphs returned 12 communities which were studied in both manual and automated manner.

**Manual Analysis of the communities:**

Small samples were taken from each of the communities and found profiles were then studied individually for characteristics which clarified if they could be members of the community or not?.

Graph shows manual sampling results

For finding out the distributions of political affiliations in different clusters, I looped through each of the clustered and looked for location of 22000 profiles whose political affiliations are known.

Graph shows the cluster distribution

| Community Number | ps | lr | fi | em | fn |
|---|---|---|---|---|---|
| Community 0 | 17 | 64 | 63 | 155 | 12 |
| Community 1 | 20 | 10 | 45 | 25 | 1 |
| Community 2 | 808 | 1200 | 130 | 2182 | 19 |
| Community 3 | 44 | 33 | 124 | 85 | 18 |

| Community 4 | 15 | 1943 | 83 | 79 | 1600 |
|-------------|-----|------|------|-----|------|
| Community 5 | 9 | 26 | 26 | 50 | 17 |
| Community 6 | 130 | 32 | 2023 | 108 | 26 |

**Discussion:**

It has been found through repeated crawls and validation by 'reference score' of nodes Twitter network that at least 3 crawls are needed to make sure that found graph will represent the network qualities of the broader graph. if seed profiles are manually annotated and large enough, it will be possible to infer political affiliations of broader network based on the known patterns of homophily in Twitter network.

Although Twitter API provides data for free but since the primary method of crawl is breadth-first approach therefore each new crawl requires exponentially more time than the previous one. To perform a large crawling exercise in Twitter, it is useful to have parallel crawlers going through nodes at the same time.

Knowing the community structure in large social networks can be useful for quantifying political polarization but one needs to make sure that there is a systematic way to factor in the margin of error that can rise from the missing nodes[5]. Data gained from conversations in Twitter can indeed provide ample proof that the communities are polarized based on the issue being discussed but the bigger question that we need to answer is that if this polarization leads to network level homophily in Twitter. This can only be done through large scale studies of evolution of a political network. Such studies will

**Conclusion:**

---

[5] (Smith, Moody et Morgan 2017)

In this Paper I have proposed a new data gathering method on Twitter keeping in mind the complications of gathering graph data and used example of French Political community structure to demonstrate the validity of the proposed method. Speed bottle-neck in the proposed method exist in so far Twitter place a restriction on the amount of data that could be gathered in a specific time window but using parallel crawling this speed can be increased considerably.